\documentclass{iopconfser}

\usepackage{a4wide,amssymb,graphicx}
\usepackage{epsfig}

\usepackage[english]{babel}
\usepackage[T1]{fontenc}
\usepackage[utf8]{inputenc}
\usepackage{amsmath}
\usepackage{amsfonts}
\usepackage{latexsym}
\usepackage{bbold}
\usepackage{calligra}
\usepackage{ulem}
\usepackage{inputenc}
\usepackage{xspace}
\usepackage{epstopdf}
\usepackage{tikz}
\usepackage{amsthm}
\usepackage{cancel}
\usepackage{comment}
\usepackage[pdftoolbar = true, pdfstartview = FitH, pdfmenubar = true]{hyperref}
\usepackage{stix}
\usepackage{relsize}

\newcommand{\w}{\mathrm{w}}
\newcommand{\weff}{\mathrm{w}_{\mathrm{eff}}}

\newcommand{\beq}{\begin{equation}}
\newcommand{\eeq}{\end{equation}}

\begin{document}

\noindent Conference Proceedings for BCVSPIN 2024: Particle Physics and Cosmology in the Himalayas\\
Kathmandu, Nepal, 9-13 December 2024 

\title{Friedmann cosmology with hyperfluids of constant equation of state}

\author{Ilaria Andrei$^{1}$\footnote{corresponding author}, Damianos Iosifidis$^{2, 3}$, Laur Järv$^{1}$ and Margus Saal$^{1}$}

\affil{$^1$Institute of Physics, University of Tartu, W.\ Ostwaldi 1, 50411 Tartu, Estonia \\ $^2$Scuola Superiore Meridionale, Largo San Marcellino 10, 80138 Napoli, Italy \\
$^3$ INFN Sezione di Napoli, Via Cintia, 80126 Napoli, Italy}

\email{ilaria.andrei@ut.ee, d.iosifidis@ssmeridionale.it, laur.jarv@ut.ee, margus.saal@ut.ee }

\begin{abstract} We discuss some aspects of cosmology in metric-affine theories of gravity where metric and affine connection are independent variables.  Such constructions, apart from the usual energy-momentum tensor, have an additional source, that of hypermomentum. Working with the cosmological principle assumption, we investigate the dynamics of the hypermomentum's degrees of freedom. In particular, we focus on the case where these degrees of freedom are proportional to the matter density and discuss the cosmological evolution depending on their associated indexes.
\end{abstract}

\section{Introduction}
The study of theories beyond Einstein's standard general relativity (GR) is constantly fed by the findings of contemporary
cosmology \cite{Perivolaropoulos:2021jda,Abdalla:2022yfr,DESI:2024mwx,CANTATA:2021asi,CosmoVerseNetwork:2025alb}. 
A way of generalizing GR is given by extending the geometric arena to that of metric-affine
gravity (MAG) \cite{Hehl:1976my, Hehl:1976kj, Hehl:1994ue}. 
In MAG, the metric is not the only fundamental geometric object, there is also an independent connection. 
This connection gives rise to two tensors, torsion and nonmetricity, which add to the curvature.
The work is done according to the cosmological principle hypothesis. The system of equations of motion involves two modified Friedmann equations along with the continuity equation for the usual matter and 
hypermomentum variables. 
The Lagrangian for the matter considered is that of a perfect hyperfluid and gives rise to isotropic hypermomentum.

The hypermomentum contributions in our case are divided into one spin and two shear components. We investigate 
the cosmological effects of these components, in addition to the usual matter energy density and pressure.

\section{Metric affine theories of gravity}

\subsection{Geometry and matter}

We consider a general 4-dimensional non-Riemannian space, endowed with a metric $g_{\mu\nu}$ and a generic affine 
connection $\Gamma^{\lambda}{}_{\mu\nu}$. 
The curvature tensor in terms of the affine connection is
\begin{equation}\label{RnonR}
    R^{\mu}_{\phantom{\mu} \nu \alpha \beta}:= 
    2\partial_{[\alpha}\Gamma^{\mu}_{\phantom{\mu}|\nu|\beta]} + 
    2\Gamma^{\mu}_{\phantom{\mu}\rho[\alpha}\Gamma^{\rho}_{\phantom{\rho}|\nu|\beta]} \,.
\end{equation}
The two additional  geometric tensors, nonmetricity and torsion, are given by
\beq
    Q_{\alpha\mu\nu}:=-\nabla_{\alpha}g_{\mu\nu}
    = \partial_{\alpha}g_{\mu\nu} + \Gamma^{\lambda}_{\phantom{\lambda}\mu\alpha}g_{\lambda\nu}
    +\Gamma^{\lambda}_{\phantom{\lambda} \nu\alpha}g_{\lambda\mu} \,,
\eeq
\beq
    S_{\mu\nu}^{\phantom{\mu\nu}\lambda}:=\Gamma^{\lambda}_{\phantom{\lambda}[\mu\nu]} 
    = \frac{1}{2}(\Gamma^{\lambda}_{\phantom{\lambda}\mu\nu} - \Gamma^{\lambda}_{\phantom{\lambda}\nu\mu})\,.
\eeq
In the matter sector, we have the usual metric energy-momentum tensor and the hypermomentum tensor 
\cite{Hehl:1976hyperm, Hehl:1978cb}, which is defined as the variation of the matter part of the action with respect 
to the independent affine connection:
\begin{align}
    T^{\alpha b} :=+\frac{2}{\sqrt{-g}}\frac{\delta(\sqrt{-g} \mathcal{L}_{M})}{\delta g_{\alpha b}} \,, \hspace{1 cm} 
    \Delta_{\lambda}^{\phantom{\lambda} \mu\nu} 
   : = -\frac{2}{\sqrt{-g}}\frac{\delta ( \sqrt{-g} \mathcal{L}_{M})}{\delta \Gamma^{\lambda}_{\phantom{\lambda}\mu\nu}} \,.
\end{align}

\subsection{Action and equation of motions}

The action we consider is the metric-affine Einstein--Hilbert term, the sole Ricci 
scalar coming from the general curvature tensor \eqref{RnonR}, along with a matter sector of a perfect hyperfluid
\cite{Iosifidis:2020gth} depending on the affine connection and on the matter fields $\chi_M$,
\begin{align}
\label{S}
    S=\frac{1}{2 \kappa}\int \mathrm{d}^{4}x \sqrt{-g} \left[ R(g_{\mu\nu}, \, \Gamma^\lambda{}_{\mu\nu}) 
    + \mathcal{L}_M \, (g_{\mu\nu}, \, \Gamma^\lambda{}_{\mu\nu}, \, \chi_M) \right] \,.
\end{align}
Varying the action \eqref{S} with respect to the metric and the connection gives us the Einstein and Palatini 
field equations \cite{Andrei_2025}
\begin{align}
    R_{(\mu\nu)}-\frac{1}{2}g_{\mu\nu}R &= \kappa T_{\mu\nu} \,, \label{metrf} \\
    \left( \frac{Q_{\lambda}}{2}+2 S_{\lambda}\right) g^{\mu\nu}-(Q_{\lambda}{}^{\mu\nu}
    +2 S_{\lambda}{}^{\mu\nu})+\left( \bar{Q}^{\mu}-\frac{Q^{\mu}}{2}-2 S^{\mu} \right)\delta^{\nu}_{\lambda} 
    &=\kappa \Delta_{\lambda}^{\phantom{\lambda} \mu\nu} \,. \label{conf}
\end{align}

We consider the cosmological principle assumption. Our metric then is the FLRW metric
\begin{align}
    \mathrm{d}s^{2}=-\mathrm{d}t^{2} + a^{2}(t)\delta_{ij} \mathrm{d}x^{i} \mathrm{d}x^{j} \,,
\end{align}
where $a(t)$ is the scale factor and measures the expansion of space. We can also define a projection tensor
\beq
\label{projop}
    h_{\mu \nu}:= g_{\mu \nu} + u_\mu u_\nu \,,
\eeq
where $u^{\mu}$ is the fluid's 4-velocity field, normalized as $u_\mu u^\mu=-1$, that in co-moving coordinates, 
is $u^\mu= \delta^\mu_0=(1,0,0,0)$.

The cosmological principle assumption greatly reduces the number of degrees of freedom of the torsion and 
nonmetricity tensors. In fact, only two components remain for the torsion \cite{tsamparlis1979cosmological}
\beq
    S_{\mu\nu\alpha}=2 u_{[\mu}h_{\nu]\alpha}\Phi(t)+\epsilon_{\mu\nu\alpha\rho}u^{\rho} P(t) 
\label{isotor}
\eeq
and three for the nonmetricity \cite{Minkevich:1998cv}
\beq 
\label{isononmet}
    Q_{\alpha \mu \nu}  = A(t) u_\alpha h_{\mu \nu} + B(t) h_{\alpha(\mu} u_{\nu)} + C(t) u_\alpha u_\mu u_\nu \,.
\eeq

As usual, the perfect fluid sector is described by the energy-momentum tensor
\beq
    T_{\mu\nu} = \rho \, u_{\mu}u_{\nu} + p \, h_{\mu\nu} \,,
\label{metrical} 
\eeq
where $h_{\mu\nu}$ the projection tensor. While for our new matter sector, the hypermomentum tensor has five components, 
$\phi$, $\chi$, $\psi$, $\omega$, $\zeta$, allowed by the cosmological principle, with covariant form \cite{Iosifidis:2020gth}
\beq 
\label{hypermomentum:degrees}
    \Delta_{\alpha\mu\nu} = \phi \, h_{\mu\alpha}u_{\nu} + \chi \, h_{\nu\alpha}u_{\mu} + \psi \, u_{\alpha}h_{\mu\nu} 
    + \omega \, u_{\alpha}u_{\mu} u_{\nu} + \epsilon_{\alpha\mu\nu\kappa}\, u^{\kappa}\zeta \,.
\eeq

Let us now discuss the equations of motion \eqref{metrf} and \eqref{conf}.
The connection equations \eqref{conf} are algebraic equations and can be used to find the 
relations between the nonmetricity, torsion, and hypermomentum variables. The metric equations \eqref{metrf} in
components are
\begin{subequations}
\label{eq: FLRW equations general}
\begin{align}
\label{eq: FR1}
    3 H^2 &= \kappa \rho + \kappa \rho_h \,, \\
\label{eq: FR2}
    2 \dot{H} + 3 H^2 &= - \kappa p - \kappa p_h \,, \\
\label{eq: continuity eq}
    \dot{\rho} + 3 H (\rho + p ) &= -\dot{\rho}_h - 3 H (\rho_h + p_h ) \,,
\end{align}
\end{subequations}
where the effective density and pressure of the hyperfluid are given by
\begin{align}\label{rhohph}
    \rho_h &:= \frac{3 \dot{\Sigma}_2}{2} + \kappa \left(- \frac{3 \Sigma_{1} \sigma}{2} + \frac{3 \Sigma_{2}^{2}}{4} 
    - \frac{3 \Sigma_{2} \sigma}{2} - \frac{3 \sigma^{2}}{4}\right) + H \left(3 \Sigma_{1} + \frac{9 \Sigma_{2}}{2} 
    + 3 \sigma\right)\,, \\
    p_h &:= \frac{\dot{\Sigma}_2}{2} - \dot{\sigma} + \kappa \left(\Sigma_{1} \Sigma_{2} - \frac{\Sigma_{1} \sigma}{2} 
    + \frac{3 \Sigma_{2}^{2}}{4} - \frac{\Sigma_{2} \sigma}{2} + \frac{\sigma^{2}}{4}\right) + H \left(\Sigma_{1} 
    + \frac{3 \Sigma_{2}}{2} - 2 \sigma\right) \,. \label{pph}
\end{align}
Note that only two of the equations \eqref{eq: FLRW equations general} are independent, since either of the last two
can be derived from the remaining two and the time derivative of the first. 
We can also define barotropic indexes of the components, as well as for the total system, as
\begin{equation}\label{w}
    \w :=\frac{p}{\rho} \,, \hspace{1 cm} \w_h := \frac{p_h}{\rho_h} \,,  \hspace{1 cm} 
    \w_{\mathrm{eff}}  := \frac{p + p_h}{\rho + \rho_h} = \frac{\w \, \rho + \w_h \, \rho_h}{\rho + \rho_h} \neq \w + \w_h \,.
\end{equation}
With our equations, it follows that
\begin{equation}\label{eq: w_eff definition}
     \w_{\mathrm{eff}} =  - 1 - \frac{2}{3} \frac{\dot{H}}{H^2} \,.
\end{equation}
These quantities are to be read in the following way: $\w$ as a barotropic index of matter, 
$\w_h$ as a hypermomentum index and $\w_{\mathrm{eff}}$ as an effective index.

The equations \eqref{eq: FLRW equations general} contain many degrees of freedom, and to be able to proceed with
their solutions and analysis we considered different possible assumptions. 
This has been done systematically in \cite{Andrei_2025}. Here we will report an example.

\section{Fluid without hypermomentum}

Here we mention the evolution equations for density and Hubble parameter in standard cosmology in the framework of general relativity but using our notation so to later make more clear the effects that arise from hypermomentum.
The density runs as
\begin{align}
\label{eq: rho(t) GR}
    \rho(t) &= \frac{\rho_0}{\left(1 \pm \frac{\sqrt{3\kappa \rho_0}}{2} (1+\w_\rho) (t-t_0) \right)^2} \,,
\end{align}
and the Hubble parameter as
\begin{align}
\label{eq: H GR)}
    H &= \frac{H_0}{1+\frac{3 H_0}{2}(1+\weff)(t-t_0)} \,.
\end{align}
In general relativity it holds that $\w = \w_\rho = \weff$. 
In the next section, we will see how this is not the case in general in metric-affine theories of gravity.

\section{Role of hypermomentum in cosmology}

Let us now analyze the modified Friedmann equations and interpret their role in cosmology. 
The systematic study of many subcases with single and multiple components has been done in the paper at the base of this
conference proceeding, \cite{Andrei_2025}. 
In the following, we deepen one of the subcases therein studied.

We choose to consider all hypermomentum quantities to be proportional to the square root of the energy density with different generic constant parameters $b$,  $b_1$, $b_2$
\begin{align}\label{sigmashear1and2}
    \sigma &= b \sqrt{\frac{3 \rho}{\kappa}} \,, \qquad \Sigma_1 = b_1 \sqrt{\frac{\rho}{3 \kappa}} \,, \qquad \Sigma_2 = b_2 \sqrt{\frac{\rho}{3 \kappa}} \,.
\end{align}
This ansatz ensures that the parameters $b$ are dimensionless. The barotropic indexes \eqref{w} in this case are no longer equal $\w =  \w_\rho = \weff$ as in general relativity. 
Considering a slow roll approximation (drop $\ddot{\rho}$ and expand for small $\dot{\rho}$) and small values for
$b$, $b_1$, $b_2$, we can calculate our indexes and the Hubble parameter at present time
to be
\begin{subequations}
\label{eq: w_rho, w_eff, H_0 in b, b_1, b2}
\begin{align}
   \w_\rho &\approx \w \pm b \mp \frac{(3\w-5)b_1}{6} \mp \frac{(3\w^2 + 20\w +1)b_2}{8} \,,\\
   \weff &\approx \w \mp \frac{(3 \w +1)b}{2} \mp \frac{(3\w-1) b_1}{3} \mp\frac{(5\w+1)b_2}{2} \,,\\
   H_0 &\approx \left( \pm 1 + \frac{3b}{2} +\frac{b_1}{2} -  \frac{3(\w-1)b_2}{8} \right) \sqrt{\frac{\kappa \rho_0}{3}} \,,\\
  \w_h &\approx +\frac{1}{6}+\frac{1}{2}\w \mp\frac{1}{8} (3 \w+1)\, b \mp \frac{2}{9 (\w-1)} b_2 \,.
\end{align}
\end{subequations}

\subsection{Equivalence with GR}

We can find a combination of the coefficients $b, b_1, b_2$ such that the hypermomentum contribution is hidden and the 
overall behavior is the same as in general relativity.
In fact, the following combination of coefficients
\begin{equation}\label{b12}
    b_1=-\frac{3\left(\w^2+6 \w+1\right)}{2 (\w+1)^2} b \,,  \hspace{2 cm} b_2=\frac{2 (\w-1)}{(\w+1)^2} b
\end{equation}
make the first order corrections of \eqref{eq: w_rho, w_eff, H_0 in b, b_1, b2} to vanish and give
\begin{equation}
 \w_\rho \approx \w \,, \hspace{1 cm} \weff \approx \w \,,  \hspace{1 cm}  H_0 = \pm \sqrt{\frac{k\, \rho_0}{3}} \,.
\end{equation}
The specific coefficients \eqref{b12} would correspond to a theory where spin, shear one and shear two are present in the form \eqref{sigmashear1and2}, but their contribution cancels out and we have the usual GR.

\subsection{Hypermomentum index divergence}

\begin{figure}[!tbp]
  \centering
  \begin{minipage}[b]{0.4\textwidth}
   \includegraphics[width=1.5\textwidth]{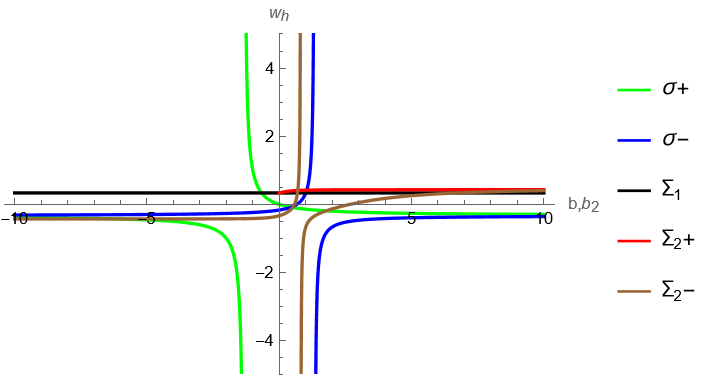}
    \caption{The $\w_h$ for $\w$ = 0}\label{figure 1}
  \end{minipage}
  \hfill
  \begin{minipage}[b]{0.4\textwidth}
 \includegraphics[width=1.3\textwidth]{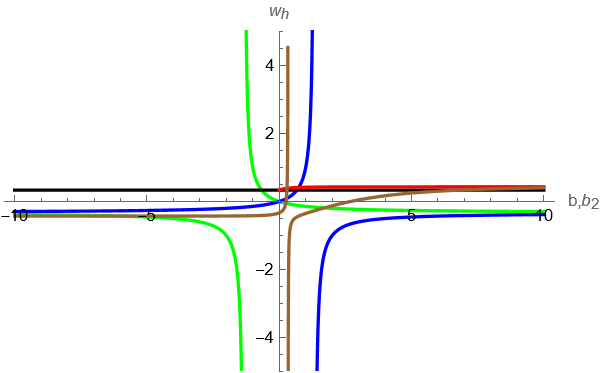}
    \caption{The $\w_h$ for $\w$ = 1/3}\label{figure 2}
  \end{minipage}
\end{figure}

Let us now reason about the hypermomentum index $\w_h$ as found directly from its definition \eqref{w} 
(without slow roll or small $b$, $b_1$, $b_2$ expansion) in order to find when it does diverge. 
In Figure \ref{figure 1} ($ \w = 0$) and Figure \ref{figure 2} ($ \w = 1/3$) the behaviour of the hypermomentum index for the three cases of spin, shear one, and shear two is represented. On the horizontal axis, there are the parameters $b$ and $b_2$ that, let us recall, 
are the coefficients in front of our chosen assumption \eqref{sigmashear1and2} for the behavior of 
$\sigma$, $\Sigma_1$ and $\Sigma_2$ as a function of the matter density $\rho$.
First, we note that $\Sigma_1$ is constant, so we have the black line at the value $1/3$ in both graphics. 
On the other hand, the cases of spin and shear two show some dangerous divergences at small $b$. This restricts 
the practical values the constant $b$ can assume in the assumption \eqref{sigmashear1and2}, and consequently limits
the validity of the expansions in \eqref{eq: w_rho, w_eff, H_0 in b, b_1, b2}.
Interestingly, we note that $\w_h$ for all the three hypermomentum quantities goes 
to $|1/3|$ for $b$, $b_1$, $b_2$ $\, \rightarrow |\infty|$ independently of the value of the matter equation of state $\w$.

\section{Conclusions}
In this proceeding, we showed how the modifications induced by the presence of torsion and nonmetricity, sourced by a hyperfluid,
change the standard cosmological scenario. We mention that we need to choose some additional assumptions about the hyperfluid
equation of state in order to be able to completely determine the dynamics of the system. We discuss one such assumption, 
namely to consider the three hypermomentum quantities $\sigma$, $\Sigma_1$, and $\Sigma_2$ proportional to the matter density $\sqrt{\rho}$ and investigate the possible values of the proportionality coefficients.
Other possible assumptions to solve the equations of motion have been considered in the paper at the basis of this proceeding \cite{Andrei_2025}.

\vspace{0.5 cm}
\bibliographystyle{utphys}
\bibliography{name}
\end{document}